\documentclass[12pt]{article}
\usepackage{UF_FRED_paper_style}
\usepackage{times}
\usepackage{amsmath, amssymb}
\usepackage{geometry}
\usepackage{cancel}
\usepackage{marginnote}
\usepackage{amsthm}

\usepackage{cleveref}
\usepackage{appendix}

\usepackage{booktabs}

\usepackage{enumitem}
\usepackage{bm}
\usepackage{graphicx}
\usepackage{booktabs}
\usepackage{url}
\usepackage{setspace}
\usepackage{enumitem}
\usepackage{subcaption}

\hypersetup{
    colorlinks = true,
    citecolor = blue
}
\usepackage{lipsum}  
\usepackage{setspace}

\title{Simulation-Free Bayesian Power and Sample Size Calculations for Bayes Factors in Single-Arm Phase II Trials with Binary Endpoints}

\author{\hspace{1.5cm}Riko Kelter\thanks{Correspondence concerning this article should be addressed to \url{rkelter@uni-koeln.de}.
    Draft version 1.0, \today. This paper has not been peer reviewed. Please do not copy or cite without author's permission. The R package \texttt{bfbin2arm} is available on CRAN, see \url{https://cran.r-project.org/web/packages/bfbin2arm/index.html}. Data and R code to reproduce all results are openly available at the Open Science Foundation under \url{https://osf.io/9fjz7/overview?view_only=6ca33c181c25425fa353dd031f5bf6a5}. The authors declare no conflict of interest.} \hspace{0.5cm} and \hspace{0.5cm} Kathrin Möllenhoff\\Institute of Medical Statistics and Computational Biology\\
	Faculty of Medicine\\
    University of Cologne\\
    Cologne, Germany\\
    }
\date{\today}

\begin{document}
\maketitle

\begin{abstract}
Bayes factors provide a coherent Bayesian measure of evidence for competing hypotheses and have recently been used as the basis for single-arm phase II trial designs with binary endpoints. In contrast to classical power analyses based on test statistics and $p$-values, Bayes-factor based sample size calculations target high probabilities of obtaining compelling evidence either for a relevant treatment effect or for the null hypothesis, given pre-specified Bayes-factor thresholds. This paper explains how to design single-arm phase II binomial trials using Bayes factors with a focus on simulation-free calibration of Bayesian and frequentist power, type-I-error, and the probability of compelling evidence for the null. Two oncology-motivated examples illustrate the approach and are implemented in the \texttt{bfbin2arm} R package, with code provided in an appendix. The methodology fits naturally into current efforts to innovate and modernize clinical trial design through Bayesian and adaptive methods.
\end{abstract}

\section{Introduction}

\subsection{Motivation}

Single-arm phase II trials with binary endpoints are widely used to decide whether a new therapy is promising enough to deserve further study \citep{Simon1989}. A typical scenario arises in oncology: patients with relapsed disease receive an experimental agent, and the primary endpoint is the objective response rate (ORR). Historical data suggest a response rate of about $20\%$ under standard care, and the trial aims to detect an improvement to around $40\%$ \citep{Chen2022}. Investigators would like the design to have a high chance to detect such an improvement while keeping the risk of falsely declaring success low.

Traditionally, these questions are answered in a frequentist framework using $p$-values and power calculations against a fixed alternative response rate \citep{Simon1989}. In contrast, Bayes factors provide an explicitly evidence-based calibration: they compare how well the data support a null hypothesis $H_0$ versus an alternative $H_1$ and update prior odds to posterior odds via multiplication by the Bayes factor \citep{KassRaftery1995,pawelHeld2025}:
\[
  \mathrm{BF}_{01}(y) 
  = \frac{f(y \mid H_0)}{f(y \mid H_1)}.
\]
Interpreted as a predictive updating factor from prior to posterior odds,
\begin{align}\label{eq:bayesFactorIntro2}
   \underbrace{\frac{P(H_0 \mid y)}{P(H_1 \mid y)}}_{\text{Posterior odds}} = 
   \underbrace{\frac{f(y \mid H_0)}{f(y \mid H_1)}}_{\text{Bayes factor }\mathrm{BF}_{01}(y)} \cdot 
   \underbrace{\frac{P(H_0)}{P(H_1)}}_{\text{Prior odds}},
\end{align}
Bayes factors separate the influence of the prior odds $P(H_0)/P(H_1)$ from the influence of the parameter priors within each hypothesis. For a fixed pair of design and analysis priors on the model parameters, the Bayes factor reflects only how the data update relative support for $H_0$ vs.\ $H_1$ \citep{VanDeSchoot2021,Bartos2022,Kelter2020BayesianPosteriorIndices,Good1983a,Kelter2022EvidenceValue}. This separation has been argued to make Bayes factors a more transparent index of evidence than posterior probabilities, especially when the prior odds on the hypotheses are themselves controversial or based on historical external information \citep{Grieve2022,Kelter2020,Linde2020,Makowski2019,Kelter2021BMCHodgesLehmann}.

Most Bayesian phase II trial designs that are currently used in practice, however, do \emph{not} calibrate operating characteristics in terms of Bayes factors and do \emph{not} admit closed-form power calculations. Instead, they typically rely on posterior probability criteria (for example, requiring that the posterior probability that the success probability $p$ of the treatment exceeds a clinically relevant threshold is large) and evaluate design performance almost exclusively via extensive Monte Carlo simulation \citep{Chevret2012,grieveIdleThoughtsWellcalibrated2016,Grieve2022,Berry2004,Berry2006,Berry2011,JohnsonCook2009,Chen2022,Lee2008,kelterBayesianGroupSequentialPredictive2024}. In this standard approach, trial outcomes are simulated under the null hypothesis $H_0$ and the alternative $H_1$ and trial operating characteristics such as the (Bayesian) power and type-I-error rate are computed as Monte Carlo averages of the simulations. If the power or type-I-error rate is too small or large, the priors or decision thresholds must be adapted and the simulation repeated. While flexible, such a simulation-based calibration has several drawbacks in routine trial planning: 
\begin{itemize}
    \item[$\blacktriangleright$]{it is computationally intensive, depending on the complexity and dimensionality of the model parameters and requires rerunning the simulation multiple times when target constraints on the operating characteristics to calibrate (such as power and type-I-error) are not met based on the chosen decision thresholds and priors,}
    \item[$\blacktriangleright$]{it introduces Monte Carlo error into quantities like Bayesian power and Bayesian type-I-error, which itself is problematic as detailed by \cite{Kelter2023,siepeSimulationStudiesMethodological2024,Boulesteix2020a,Boulesteix2020,Boulesteix2018,Boulesteix2017,Seibold2021,Morris2019}. Decreasing the Monte Carlo standard error requires increasing the number of simulations, which in turn aggravates the first problem.}
    \item[$\blacktriangleright$]{it can make it difficult to verify whether constraints are truly satisfied across many design scenarios \citep{KelterPawel2025}, as the number of simulations quickly becomes large, in turn contributing to the first and second problem.} 
\end{itemize}
These limitations become particularly acute when trialists wish to explore multiple priors, evidence thresholds, or additional criteria such as the probability of strong evidence for the null hypothesis. Importantly, next to the computational issues like long runtimes and the uncertainty introduced by Monte Carlo error,
\begin{itemize}
    \item[$\blacktriangleright$]{communicating and reporting a Bayesian trial design becomes inherently difficult, as regulatory agencies and collaborators need to understand and recreate \textit{all} simulation results. In a recent analysis, \cite{marksSystematicReviewSample2026} therefore found that Bayesian methods for sample size determination are rarely used in trial design. The main reason is a lack of transparency in reporting due to the simulation-based nature of Bayesian sample size determination (including power and type-I-error calibration), despite significant theoretical development.}
\end{itemize}

\subsection{Aims and scope}

Significant advances have recently been made in the literature on Bayesian sample size determination and design calibration \citep{KelterPawel2025,pawelHeld2025,pawelBayesFactorGroup2026,kelterTwoArmTwoStage2026,liBayesianDesignNonInferiority2018,KelterPawelTwoStage2025}. This includes, in particular, an entirely simulation-free approach to calibrating and designing a Bayesian phase II trial with binary endpoints, which is outlined in \Cref{sec:simfree}.

The goal of this paper is to translate this theoretical methodological work into a practical overview for single-arm one-stage designs with binary endpoints based on Bayes factors. It provides a conceptual bridge between the theoretical background and practical challenges when being faced with the task of calibrating a Bayesian phase II trial, and aims at clinicians and statisticians interested in calibrating Bayesian phase II trials in a practical setting. The focus is on
\begin{itemize}
    \item[(i)]{explaining how to formulate Bayes-factor decision rules in clinically meaningful terms}
    \item[(ii)]{showing how to calibrate Bayesian power, Bayesian and frequentist type-I-error, and the probability of compelling evidence for the null without simulation, and}
    \item[(iii)]{illustrating the approach in two oncology-motivated phase II trials, which serve as practical examples, using the \texttt{bfbin2arm} R package \citep{kelterTwoArmTwoStage2026}. The latter R software package implements all of the abovementioned calibration routines and allows for simple application of the methodology.}
\end{itemize}

To make ideas concrete, we use a realistic single-arm oncology scenario in this paper similar to those considered in recent Bayesian two-stage designs \citep{Chen2022}. Let $Y$ denote the number of responses among $n$ treated patients, $Y \sim \mathrm{Binomial}(n,p)$, where $p$ is the unknown objective response rate (ORR). The design problem we focus on in this paper is to compare:
\begin{itemize}
  \item a null hypothesis $H_0$ representing an unacceptable response rate, e.g.\ $p \le p_0=0.2$,
  \item an alternative $H_1$ representing a clinically relevant improvement, e.g.\ $p>p_0$ with most prior mass near $p_1\approx 0.4$.
\end{itemize}
In Section~\ref{sec:examples} below, two complete designs for such a setting are developed using Bayes factors via the R software package \texttt{bfbin2arm}: the first matches targets on power and type-I-error only, while the second additionally controls the probability of compelling evidence for the null. 

\subsection{Simulation-free Bayesian sample size determination and design calibration}
\label{sec:simfree}
\begin{figure}[!htb]
    \centering
    \includegraphics[width=1.0\linewidth]{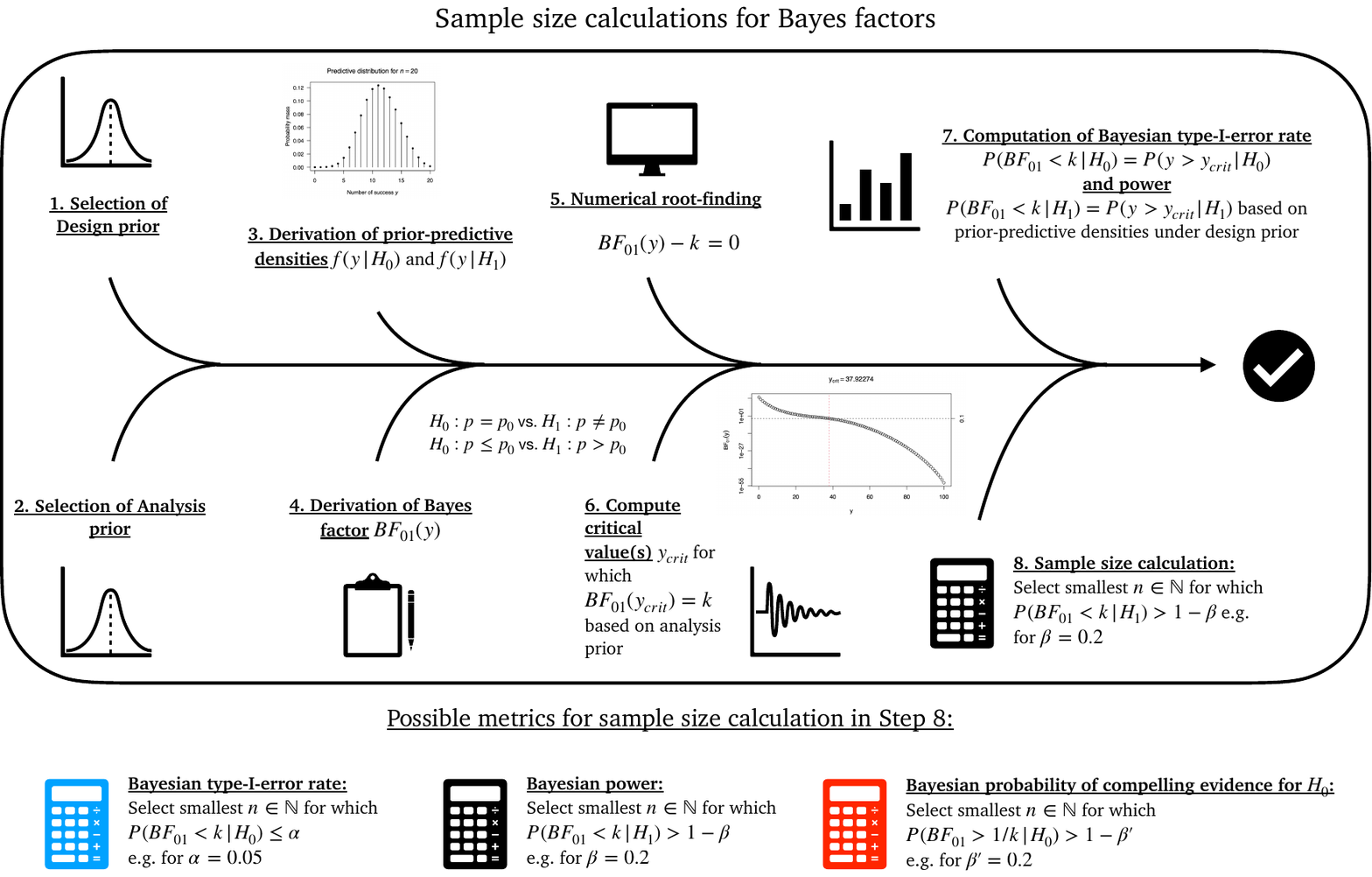}
    \caption{Overview of Bayesian power and sample size calculations for the case of a single-arm phase II trial with a binary endpoint, using Bayes factors. Details are provided in \cite{KelterPawel2025}.}
    \label{fig:flowchart_rootfinding}
\end{figure}

In this section, we provide a brief overview about the methodological advances which have recently been made towards a simulation-free power and sample size determination for Bayesian trials. In recent methodological work, \cite{KelterPawel2025} proposed a methodological shift in Bayesian sample size determination away from Monte Carlo simulations towards a numerical approach. The key idea is to replace the simulation of trial outcomes under $H_0$ and $H_1$ and reporting operating characteristics such as the (Bayesian) power and type-I-error as Monte Carlo averages by a root-finding algorithm. This algorithm is visualized in \Cref{fig:flowchart_rootfinding}, which visualizes the process of Bayesian power and sample size calculations for the single-arm phase II trial case with a binary endpoint, compare \cite{KelterPawel2025}, and involves eight steps.

\begin{figure}[!htb]
    \centering
    \includegraphics[width=0.8\linewidth]{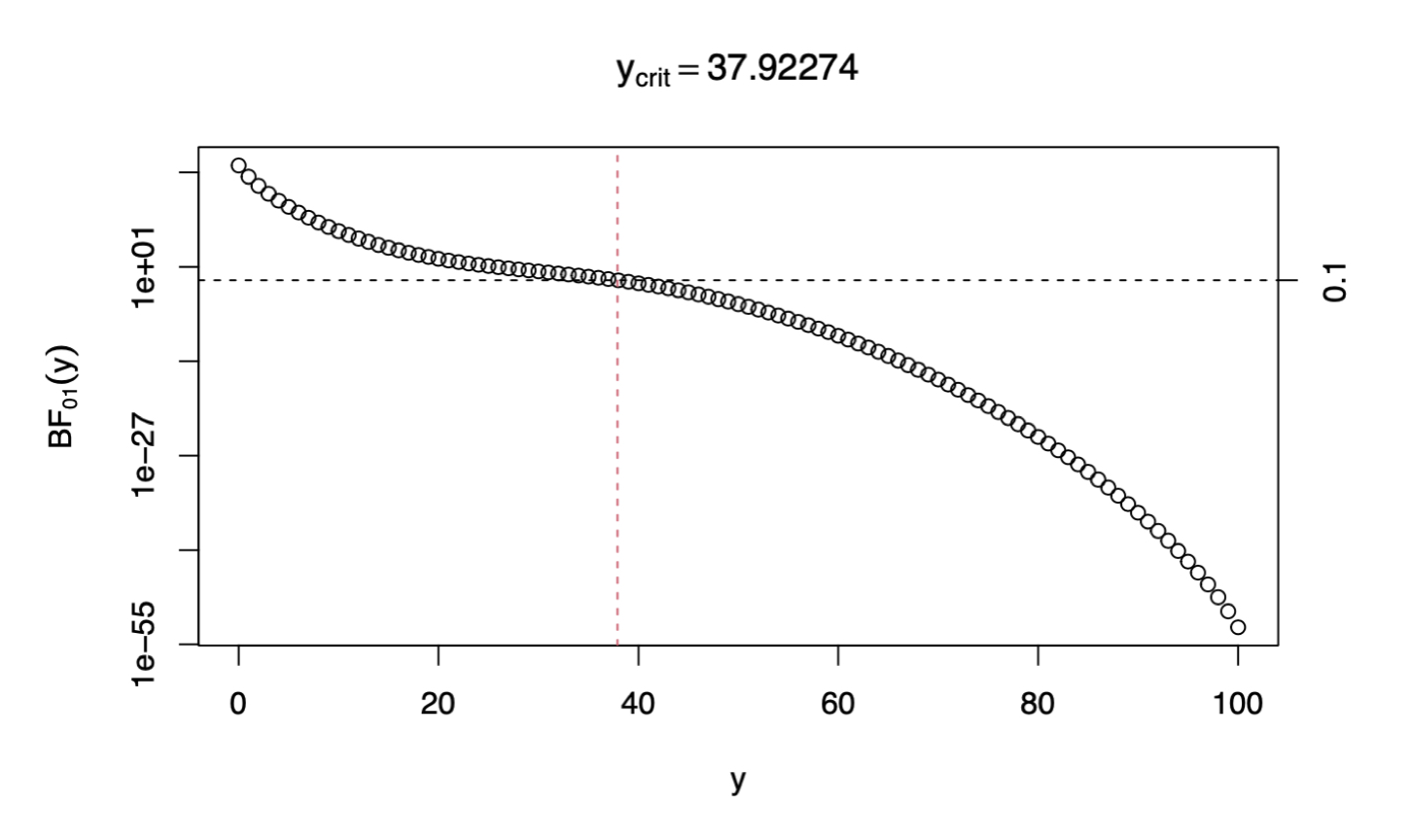}
    \caption{Relationship between the number of successes $y$ and the Bayes factor $BF_{01}(y)$ for the case of a single-arm phase II trial with a binary endpoint. The calibration algorithm searches between the sample sizes $n=0$ and $n=100$ and finds the solution $y_{crit}=37.92274$ for the equation $BF_{01}(y)-k=0$ for $k=1/10$. Details are provided in \cite{KelterPawel2025}.}
    \label{fig:ycritBFrelationship}
\end{figure}

The first step includes the selection of the design and analysis priors. Design priors are the priors chosen on the model parameters which reflect our expectation about the treatment effect. These priors influence the operating characteristics of the resulting trial design like the power and type-I-error rate, but do not influence the Bayes factor calculated during the actual trial when the trial data are available. The latter is based on the analysis prior, which is the prior used in the computation of the Bayes factor itself. \cite{KelterPawel2025} derived Bayes factors and the prior-predictive densities $f(y|H_0)$ and $f(y|H_1)$ (step 3. and 4. in \Cref{fig:flowchart_rootfinding}) for point-null and composite hypotheses in the binomial setting and showed that Bayesian power, Bayesian type-I-error, and related operating characteristics can be computed numerically without simulation in both single-arm and two-arm binomial phase II designs \citep{KelterPawel2025,kelterTwoArmTwoStage2026,KelterPawelTwoStage2025}. Therefore, they proposed a root-finding algorithm which solves the equation
$$BF_{01}(y)-k=0$$
and thereby isolates the critical threshold $y_{crit}$ based on which the relationship
$$BF_{01}(y)<k \Leftrightarrow y>y_{crit}$$
holds. In practical terms, the Bayes factor crosses the evidence threshold $k$ if and only if there are at least $y_{crit}$ treatment successes in the phase II trial. \Cref{fig:ycritBFrelationship} shows this relationship for a phase II trial where the 
 precise value of $y_{crit}$ clearly depends on the design priors, the analysis priors, and the selected evidence threshold $k$. After this fifth and sixth step -- see \Cref{fig:flowchart_rootfinding} -- the last two steps include the computation of the Bayesian type-I-error rate
\begin{align}
    P(BF_{01}<k|H_0)=P(y>y_{crit}|H_0)
\end{align}
and Bayesian power
\begin{align}
    P(BF_{01}<k|H_1)=P(y>y_{crit}|H_1)
\end{align}
based on the prior-predictive densities $f(y|H_0)$ and $f(y|H_1)$ (compare step 3.). Sample size determination then simply proceeds by picking the smallest $n \in \mathbb{N}$ for which the (Bayesian) power $P(BF_{01}(y)<k|H_1)>1-\beta$ for a prespecified $\beta \in (0,1)$, e.g. $\beta = 0.2$.

Possible metrics to calibrate the one-stage single-arm phase II trial design are shown in the bottom panel of \Cref{fig:flowchart_rootfinding}: These include the Bayesian type-I-error rate, the Bayesian power, and the probability of compelling evidence for the null hypothesis $H_0$. We provide details on these metrics later.

In the single-arm one-stage case, the approach of \cite{KelterPawel2025} leads to a \emph{simulation-free} Bayesian analogue of classical power analysis: for given priors, evidence thresholds, and design targets, the required sample size can be obtained by simple one-dimensional numerical search rather than Monte Carlo simulation \citep{KelterPawel2025}. Extensions to two-stage designs and optimal stopping rules for single-arm trials have also been developed within the same framework \citep{KelterPawelTwoStage2025}, but these are beyond the scope of the present tutorial. We only briefly outline the existing methodological work in \Cref{tab:1}:
\begin{table}[t]
  \centering
  \caption{Extensions of the Bayes factor design framework}
  \label{tab:1}
  \begin{tabular}{p{0.2\textwidth} p{0.2\textwidth} p{0.55\textwidth}}
    \hline
    Reference & Design type & Key contribution \\
    \hline
    \cite{KelterPawel2025} & Single-arm, one-stage & Includes the basic single-arm one-stage design which allows simulation-free Bayesian power and sample size calculations for binary endpoints.\\
    \cite{KelterPawelTwoStage2025} 
      & Single-arm, two-stage (sequential)
      & Extended the approach to sequential two-stage designs by introducing an interim analysis that allows early stopping for futility, leading to optimal Bayesian trial designs that minimize the expected sample size under $H_0$. \\[0.75ex]
    \cite{kelterTwoArmTwoStage2026} 
      & Two-arm, one-stage
      & Extended the approach to one-stage two-arm designs, enabling calibration of phase~II trials with binary endpoints in both treatment and control arms. \\[0.75ex]
    \cite{kelterOptimalSequentialTwostage2026} 
      & Two-arm, two-stage (sequential)
      & Extended the approach to sequential two-stage two-arm designs by introducing an interim analysis that allows early stopping for futility, yielding optimal Bayesian two-arm designs that minimize the expected sample size under $H_0$. \\
    \hline
  \end{tabular}
\end{table}
Next to these available extensions, calibration approaches for Bayesian equivalence tests and flexible extensions that allow early stopping for efficacy in addition to early stopping for futility -- for an outline see \cite{kelterOptimalSequentialTwostage2026} -- are currently being developed.

\subsection{Outline}

\section{Bayes factors as evidence}\label{eq:bf}

The formal definition of Bayes factors for point-null and composite hypotheses in the binomial setting, together with closed-form expressions under beta priors, is given in the accompanying methodological paper \citep{KelterPawel2025}. Here, only the main ideas are summarized.

\subsection{Predictive evidence comparison}

Given two hypotheses $H_0$ and $H_1$ with priors on $p$, the Bayes factor is
\[
\mathrm{BF}_{01}(y) = \frac{P(Y=y \mid H_0)}{P(Y=y \mid H_1)},
\]
where $P(Y=y \mid H_j)$ is the predictive probability of observing $Y=y$ under $H_j$, obtained by integrating the binomial likelihood over the prior on $p$ \citep{KassRaftery1995,pawelHeld2025,KelterPawel2025}. Values $\mathrm{BF}_{01} > 1$ support $H_0$, whereas $\mathrm{BF}_{01} < 1$ support $H_1$.

Thresholds such as $\mathrm{BF}_{01}<1/3$ or $\mathrm{BF}_{01}<1/10$ are often used to define what constitutes moderate or strong evidence against $H_0$ \citep{KassRaftery1995}. In a design context, these thresholds are translated into decision rules and calibration targets.

\subsection{Composite hypotheses and priors}

In single-arm phase II oncology, it is natural to treat any ORR at or below $p_0$ as clinically uninteresting \citep{Chen2022}. We therefore use composite hypotheses
\[
H_0: p \le p_0, \qquad H_1: p>p_0,
\]
with priors $\pi_0(p)$ on $[0,p_0]$ and $\pi_1(p)$ on $(p_0,1]$. A convenient choice, used throughout the examples, is a truncated $\mathrm{Beta}(1,1)$ prior (uniform) on each interval \citep{KelterPawel2025}. The resulting predictive distributions under $H_0$ and $H_1$ can be evaluated in closed form, and the corresponding Bayes factors are implemented in \texttt{bfbin2arm} \citep{KelterPawel2025}.

\subsection{Evidence thresholds and decision regions}

We distinguish two thresholds:
\begin{itemize}
  \item $k$: evidence threshold for efficacy, declaring the treatment promising if $\mathrm{BF}_{01}<k$;
  \item $k_f$: evidence threshold for futility, declaring the treatment non-promising if $\mathrm{BF}_{01}>k_f$.
\end{itemize}
The region $k \le \mathrm{BF}_{01} \le k_f$ corresponds to inconclusive evidence. In the examples, we use values $k=1/10$ or $k=1/3$, corresponding to strong or moderate evidence in favour of $H_1$ \citep{KassRaftery1995}. Also, we use $k_f=3$ in most cases, because strong evidential requirements on the probability of compelling evidence often lead to very large sample sizes.

\section{Design metrics}\label{eq:metrics}

\subsection{Bayesian and frequentist power}

Bayesian power for efficacy,
\[
\pi_{\mathrm{B}}^{(1)}(n; k_1) = P(\mathrm{BF}_{01}(Y) < k \mid H_1),
\]
is the predictive probability under $H_1$ that the trial yields at least a Bayes factor $k$ in favour of efficacy \citep{KelterPawel2025}. A frequentist counterpart conditions on a fixed $p_1$ and computes
\[
\pi_{\mathrm{F}}^{(1)}(n; k_1, p_1) = P(\mathrm{BF}_{01}(Y) < k \mid p=p_1).
\]
Both quantities can be evaluated without simulation by summing binomial or predictive probabilities over the response counts $y$ that fall into the evidence region $\{\mathrm{BF}_{01}(y)<k\}$ \citep{KelterPawel2025}. In terms of \Cref{fig:ycritBFrelationship}, these are precisely the response counts $y>y_{crit}=37.92274$. Thus, the Bayesian power in that case is the predictive probability of obtaining $y=38$ or more successes out of $n=100$. The corresponding calculations are wrapped in the single-arm functions of \texttt{bfbin2arm}.

\subsection{Bayesian and frequentist type-I-error}

Bayesian type-I-error is
\[
\alpha_{\mathrm{B}}(n; k_1) = P(\mathrm{BF}_{01}(Y) < k \mid H_0),
\]
the predictive probability under $H_0$ of falsely obtaining compelling evidence for efficacy \citep{KelterPawel2025}. The frequentist analogue at $p_0$ is
\[
\alpha_{\mathrm{F}}(n; k_1, p_0) = P(\mathrm{BF}_{01}(Y) < k \mid p=p_0).
\]
These quantities play the same conceptual role as type-I-error in classical designs but are expressed in terms of evidence thresholds.

\subsection{Probability of compelling evidence for the null}

The probability of compelling evidence for the null is defined in terms of the event
\[
\mathrm{CE}(k_0) = \{y\in \mathcal{Y} \mid \mathrm{BF}_{01}(y)>k_0\},
\]
that is, at least evidence factor $k_0$ in favour of $H_0$ \citep{KelterPawel2025}. Two predictive probabilities are of special interest:
\begin{itemize}
  \item $P(\mathrm{CE}(k_0)\mid H_0)$, the probability that an ineffective treatment is recognized as such with compelling evidence,
  \item $P(\mathrm{CE}(k_0)\mid H_1)$, the probability that an effective treatment is mistakenly supported by compelling evidence for $H_0$.
\end{itemize}
These probabilities quantify how well a design “kills bad drugs early” and how rarely it discards truly effective treatments with strong evidence \citep{kelterOptimalSequentialTwostage2026}. More important in practical settings is the first probability, however. The examples below show how to include a lower bound on $P(\mathrm{CE}(k_0)\mid H_0)$ in the calibration.

\section{Worked examples using \texttt{bfbin2arm}}
\label{sec:examples}

This section presents the analysis of two clinical trials designs reported in the literature on oncology. The first trial is a single-arm phase II trial with binary endpoints investigating the efficacy of a novel treatment for gastric or gastroesophageal junction cancer which was published in \textit{The Lancet Oncology} recently \citep{cutsemTrastuzumabDeruxtecanPatients2023}. The second trial is a single-arm phase II trial with binary endpoints studying the efficacy of a novel treatment against non-small cell lung cancer (NSCLC) and was recently published in \textit{Nature Medicine} \citep{chaftNeoadjuvantAtezolizumabResectable2022}. In the subsections below, we illustrate how to design and calibrate a Bayesian phase II trial in the context of these oncology settings. In particular, we demonstrate how to elicit the design priors, choose evidence thresholds, and calibrate a trial using Bayesian, frequentist or hybrid criteria. R implementation details are given in Appendix~\ref{app:rcode}.

\subsection{DESTINY-Gastric02: second-line T-DXd in HER2-positive gastric cancer}

A recent and clinically influential single-arm phase II trial is DESTINY-Gastric02, which evaluated trastuzumab deruxtecan (T-DXd) in Western patients with unresectable or metastatic HER2-positive gastric or gastroesophageal junction cancer whose disease had progressed on or after a trastuzumab-containing regimen \citep{cutsemTrastuzumabDeruxtecanPatients2023}. The primary endpoint was confirmed objective response rate (ORR) by independent central review. Historical second-line data from the RAINBOW trial suggested that standard chemotherapy produced an ORR of about $27\%$ \citep{cascinuTumorResponseSymptom2021}, which was taken as the benchmark rate $p_0$ for an uninteresting response \citep{cutsemTrastuzumabDeruxtecanPatients2023,sidawayTrastuzumabDeruxtecanImproves2020}. In contrast, T-DXd was expected to achieve an ORR around $45\%$, denoted $p_1$, based on DESTINY-Gastric01 and early phase data \citep{kangTrastuzumabDeruxtecanReview2023,cutsemTrastuzumabDeruxtecanPatients2023}.

In the original frequentist design, the null hypothesis was that the true ORR did not exceed $27\%$, and the study was powered to show that the lower bound of the $95\%$ confidence interval for ORR would be above $27\%$ with approximately $90\%$ power for $n=72$ evaluable patients \citep{cutsemTrastuzumabDeruxtecanPatients2023}. Translated into a Bayes-factor framework, a natural formulation is a composite null
\[
  H_0: p \leq 0.27
\]
representing clinically uninteresting ORR at or below the historical benchmark, and an alternative
\[
  H_1: p > 0.27,
\]
with a design prior on $p$ under $H_1$ centred at $p_1 = 0.45$ to reflect the anticipated improvement. For example, under $H_0$ one may use a truncated beta prior on $[0,0.27]$ with mean close to $0.20$, while under $H_1$ a truncated beta prior on $(0.27,1]$ with mean $0.45$ concentrates mass around the expected T-DXd effect. This preserves the clinical interpretation that any ORR up to about $27\%$ is uninteresting, whereas larger values are considered promising, and links directly to the original sample size justification. Table~\ref{tab:destiny_gastric02} summarizes the main design quantities for DESTINY-Gastric02.

\begin{table}[ht]
\centering
\caption{Key design quantities for DESTINY-Gastric02 (single-arm phase II T-DXd trial), compare \cite{cutsemTrastuzumabDeruxtecanPatients2023}.}
\label{tab:destiny_gastric02}
\begin{tabular}{ll}
\toprule
Quantity & Value \\
\midrule
Indication & HER2-positive advanced gastric / GEJ cancer \\
Design & Single-arm phase II, Western patients (USA + Europe) \\
Primary endpoint & Confirmed ORR (binary) by independent central review\\
Benchmark rate $p_0$ & 0.27 (historical second-line ORR from RAINBOW)\\
Target rate $p_1$ & 0.45 (anticipated ORR under T-DXd)\\
Null hypothesis & $H_0: p \leq 0.27$ \\
Alternative hypothesis & $H_1: p > 0.27$ (design prior centred at $p_1=0.45$) \\
Planned sample size & 72 evaluable patients (79 treated)\\
Power target & 90\% (lower 95\% CI bound for ORR above 0.27)\\
Observed ORR & 38\% (primary analysis), 42\% (updated; 95\% CI 30.8--53.4)\\
\bottomrule
\end{tabular}
\end{table}

In the Bayes-factor framework, we declare efficacy when \(BF_{01} \le k\) and compelling evidence for \(H_0\) when \(BF_{01} \ge k_{f}\), with moderate evidence thresholds \(k = 0.1\) and \(k_{f} = 3\) in Jeffreys' orientation \citep{jeffreys1961}. Using the \texttt{bfbin2arm} package\footnote{See \url{https://cran.r-project.org/web/packages/bfbin2arm/index.html}. Further information about the package is available at \url{https://imsb.uni-koeln.de/en/research/research-highlights/bfbin2arm}.}, we calibrated single-arm one-stage designs under three modes---Bayesian, frequentist, and hybrid---and examined how different design priors affect the resulting sample size.

\subsubsection{Bayesian calibration with a diffuse design prior}

\begin{figure}[h!]
  \centering
  \includegraphics[width=\textwidth]{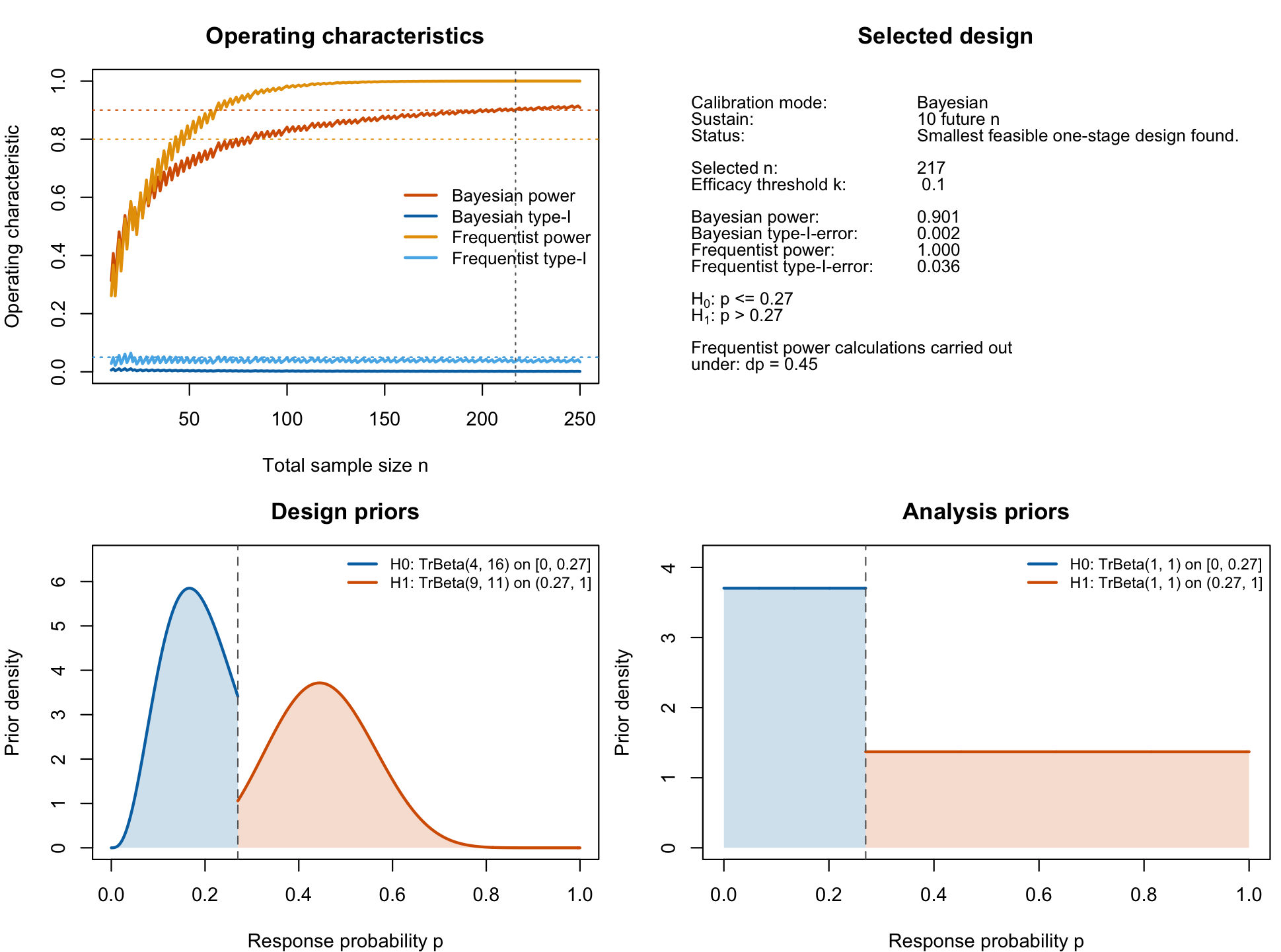}
  \caption{Bayesian calibration for DESTINY-Gastric02 with baseline design priors under \(H_0\) and \(H_1\). Bottom panels: truncated Beta design and analysis priors; top left panel: Bayes-factor operating characteristics (Bayesian power, type-I error, and \(\mathrm{PCE}(H_0)\)) as functions of the sample size \(n\); top right panel: Key summary of the resulting calibrated trial design.}
  \label{fig:destiny-bayes-baseline}
\end{figure}

We first chose design priors that concentrate mass near the clinically relevant response rates. Under \(H_0\), the design prior is a truncated \(\mathrm{Beta}(4, 16)\) on \([0, p_0]\), with mean \(4/(4+16) \approx 0.20\) before truncation, reflecting the belief that responses above \(27\%\) are unlikely when the novel drug is ineffective. Under \(H_1\), a baseline design prior is a truncated \(\mathrm{Beta}(9, 11)\) on \((p_0, 1]\), with mean \(9/(9+11) \approx 0.45\), centred around the anticipated ORR under T-DXd. These priors are shown in the bottom left panel of Figure~\ref{fig:destiny-bayes-baseline}, and the corresponding flat analysis priors under $H_0$ and $H_1$ are shown in the bottom right panel. The top right panel in Figure~\ref{fig:destiny-bayes-baseline} shows the resulting Bayes-factor design under pure Bayesian calibration with these priors, requiring a sample size of \(n = 217\) patients. The Bayesian operating characteristics at this \(n\) are: Bayesian power \(0.901\) and Bayesian type-I error \(0.002\), while the frequentist power at \(p_1 = 0.45\) is \(1.000\) and the frequentist type-I error at \(p_0 = 0.27\) is \(0.036\). The top left panel in Figure~\ref{fig:destiny-bayes-baseline} shows the Bayes-factor operating characteristics as functions of the sample size.

\begin{figure}[h]
  \centering
  \includegraphics[width=\textwidth]{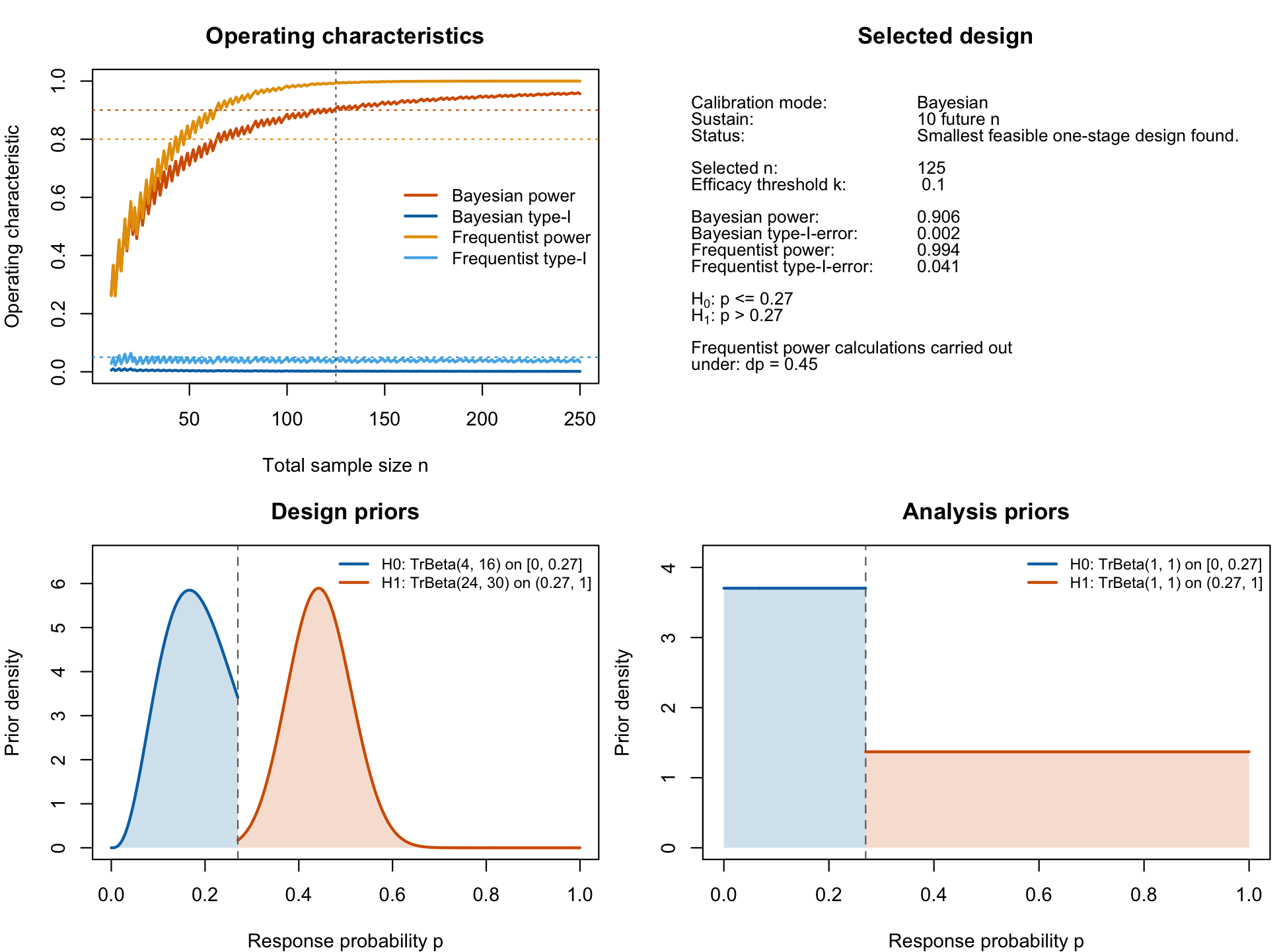}
  \caption{Bayesian calibration for DESTINY-Gastric02 with a more concentrated design prior under \(H_1\) (truncated \(\mathrm{Beta}(24, 30)\)), keeping the \(H_0\) prior unchanged. The increased concentration around \(p_1 = 0.45\) allows the Bayesian evidence criteria to be met with a smaller sample size (\(n = 125\)).}
  \label{fig:destiny-bayes-conc}
\end{figure}

\subsubsection{Bayesian calibration with a more concentrated design prior}

To illustrate the impact of the design priors under \(H_1\), we then replaced the baseline truncated \(\mathrm{Beta}(9, 11)\) prior by a more concentrated truncated \(\mathrm{Beta}(24, 30)\) prior, which has roughly the same prior mean $24/(24+30) \approx 0.44$ but a larger concentration around \(p_1\). This prior is shown in the bottom left panel in \Cref{fig:destiny-bayes-conc}. Under otherwise identical settings and thresholds, the Bayesian calibration now yields a much smaller sample size of \(n = 125\), with Bayesian power \(0.906\) and Bayesian type-I error \(0.002\). Frequentist power and type-I error at \(p_1\) and \(p_0\) are \(0.994\) and \(0.041\), respectively. The comparison in Figure~\ref{fig:destiny-bayes-conc} makes the effect of the design prior transparent: a more concentrated \(H_1\) design prior effectively separates the hypotheses and places more prior mass on response probabilities around \(0.45\), which allows the Bayes-factor calibration algorithm to reach the power and \(\mathrm{PCE}(H_0)\) targets with a substantially smaller \(n\) than under the more diffuse baseline prior. In other words, increasing prior informativeness under \(H_1\) translates directly into a reduction of the Bayesianly calibrated sample size.

\subsubsection{Adding a constraint on the probability of compelling evidence}

To connect this Bayesian design to long-run frequentist criteria and to demonstrate the role of \(\mathrm{PCE}(H_0)\) constraints, we added a minimum prior-predictive probability of compelling evidence under \(H_0\).

\begin{figure}[h]
  \centering
  \includegraphics[width=\textwidth]{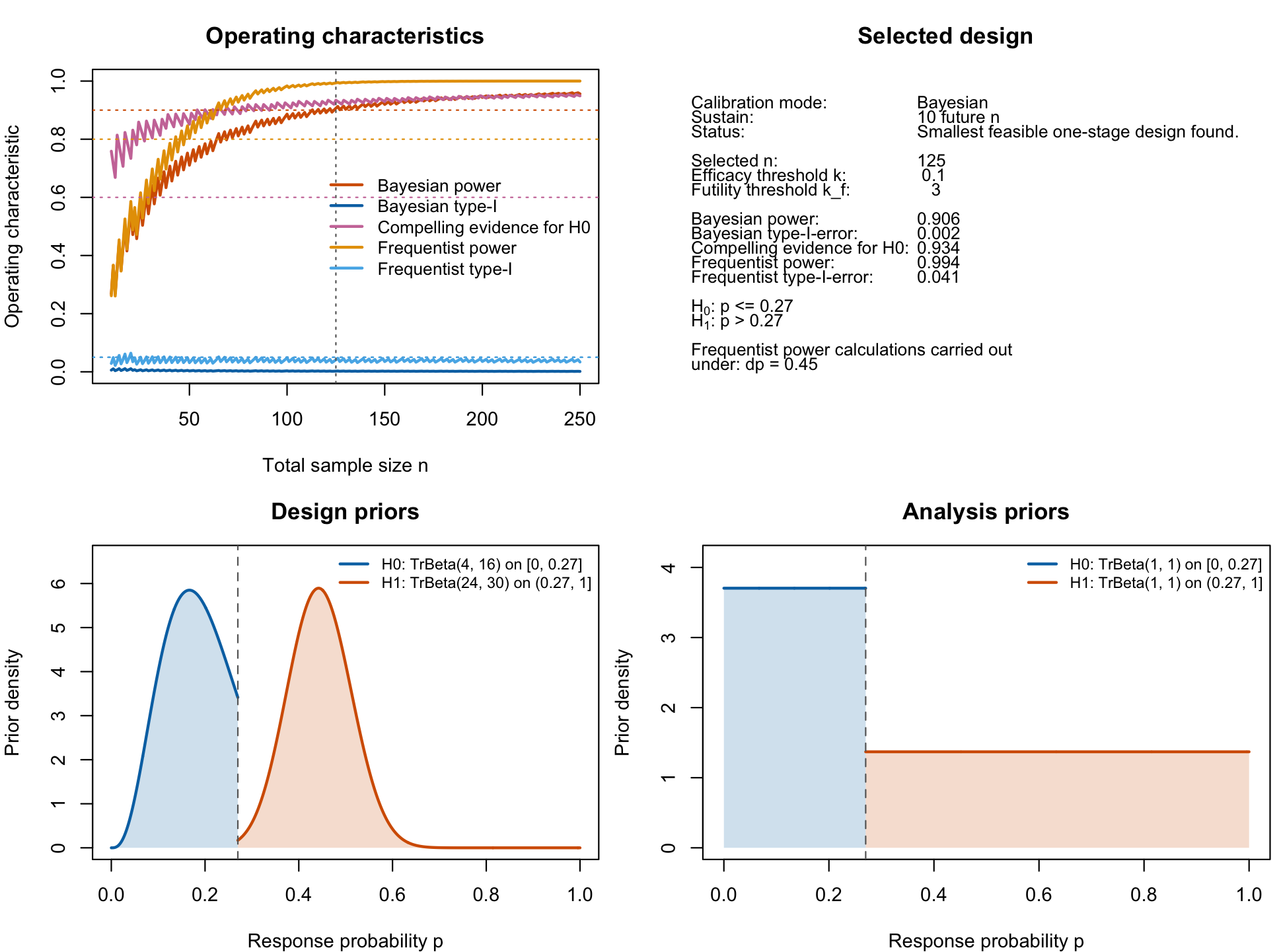}
  \caption{Bayesian calibration for DESTINY-Gastric02 with a more concentrated design prior under \(H_1\) (truncated \(\mathrm{Beta}(24, 30)\)), keeping the \(H_0\) prior unchanged. The increased concentration around \(p_1 = 0.45\) allows the Bayesian evidence criteria to be met with a smaller sample size (\(n = 125\)). In contrast to \Cref{fig:destiny-bayes-conc}, now a constraint of 60\% on the probability of compelling evidence $\mathrm{PCE}(H_0)$ for $H_0$ is added.}
  \label{fig:destiny-bayes-conc-with-ce}
\end{figure}

With \(k_f = 3\) and \(\mathrm{PCE}(H_0) \ge 0.60\), the concentrated prior design still selects \(n = 125\) and \(\mathrm{PCE}(H_0) = 0.934\), showing that the CE constraint is non-binding at this moderate level, compare \Cref{fig:destiny-bayes-conc-with-ce}. The reason can be seen in the top left panel in \Cref{fig:destiny-bayes-conc-with-ce}, as the probability of compelling evidence is not the limiting factor here. The Bayesian power achieves its target constraint of 90\% only for $n=125$, at which point the other operating characteristics have already passed their respective thresholds.

\subsubsection{Frequentist calibration}

So far, the calibration mode was Bayesian and thus, Bayesian power and type-I-error were the key operating characteristics according to which the trial sample size was selected. Alternatives to a Bayesian calibration are frequentist, hybrid and full calibration modes. We illustrate frequentist calibration first, where instead of Bayesian power and type-I-error, the corresponding frequentist quantities are the target constraints for calibration. Thus, now 90\% frequentist power and 5\% frequentist type-I-error are the target constraints.

\begin{figure}[h]
  \centering
  \includegraphics[width=\textwidth]{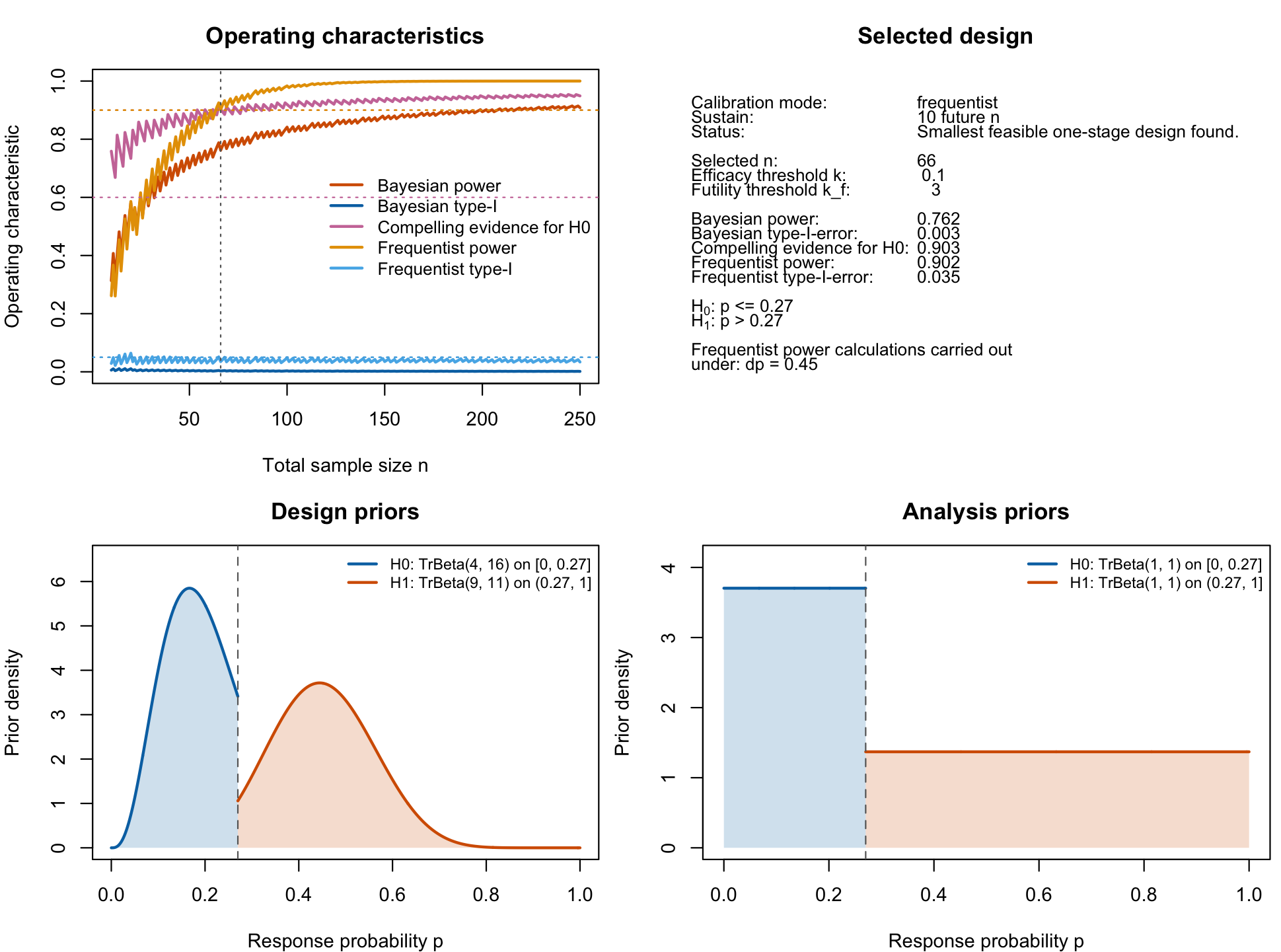}
  \caption{Frequentist calibration for DESTINY-Gastric02 using the baseline design priors. The design enforces frequentist power \(0.90\) at \(p_1 = 0.45\) and frequentist type-I error \(0.05\) at \(p_0 = 0.27\), while maintaining \(\mathrm{PCE}(H_0) \ge 0.60\). The resulting sample size is \(n = 66\).}
  \label{fig:destiny-freq-with-ce}
\end{figure}

In frequentist calibration mode, the design selects a much smaller sample size \(n = 66\), with \(\mathrm{PCE}(H_0) = 0.903\) and frequentist power and type-I error at \(p_1\) and \(p_0\) equal to \(0.90\) and \(0.05\), respectively. Figure~\ref{fig:destiny-freq-with-ce} shows that frequentist calibration, which prioritises long-run error control at fixed \(p_0\) and \(p_1\), can yield a more aggressive sample size than the Bayesian calibration, even in the presence of a \(\mathrm{PCE}(H_0)\) constraint. A further point worth emphasizing is that a target constraint on the probability of compelling evidence for $H_0$ can be added to a fully frequentist calibration. This makes a fully frequentist calibration attractive from a frequentist point of view due to two reasons:
\begin{itemize}
    \item[$\blacktriangleright$]{First, the Bayes factor allows to express evidence in favour of the null hypothesis $H_0$ and the alternative $H_1$. Strictly frequentist designs which make use of p-values cannot express evidence in favour of $H_0$, even if the p-value is large. If the significance threshold is not passed, no statement about the validity of the null hypothesis $H_0$ can be made via a p-value.}
    \item[$\blacktriangleright$]{Second, and relating to the first point, not only can evidence about $H_0$ be expressed by the Bayes factor. In addition, the calibration of \(\mathrm{PCE}(H_0)>0.60\) ensures that if $H_0$ indeed is true, sufficient evidence $k_f$ in form of a Bayes factor $BF_{01}>k_f$ will be found at least with probability $60\%$. This allows to identify an ineffective drug if it indeed does not surpass the required benchmark success rate $p_0$.}
\end{itemize}
Both points above hold, even if the Bayes factor is merely a frequentist test statistic when calibrating a design in the frequentist calibration mode.

\begin{figure}[ht]
  \centering
  \includegraphics[width=\textwidth]{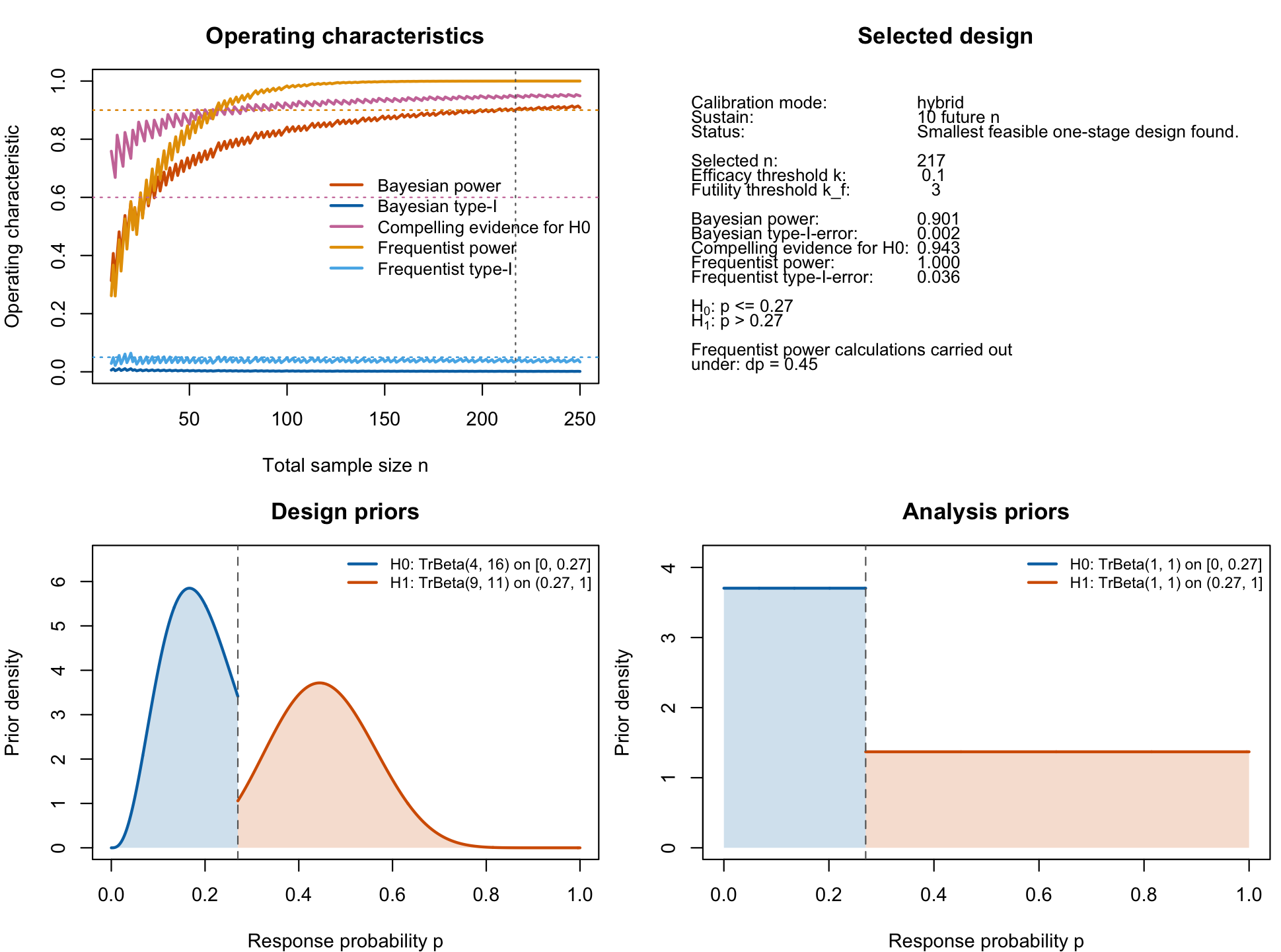}
  \caption{Hybrid calibration for DESTINY-Gastric02 with baseline design priors. The design simultaneously enforces Bayesian power and frequentist type-I error constraints, leading to a sample size \(n = 217\) and \(\mathrm{PCE}(H_0) = 0.943\), matching the pure Bayesian baseline design in this case.}
  \label{fig:destiny-hybrid-with-ce}
\end{figure}

\subsubsection{Hybrid calibration: Combining frequentist and Bayesian criteria}

Finally, the hybrid calibration, which enforces 80\% Bayesian power while constraining frequentist $\leq 5\%$ type-I error, returns a sample size of \(n = 217\) with \(\mathrm{PCE}(H_0) = 0.943\), identical to the pure Bayesian baseline design in this example. The corresponding plot in Figure~\ref{fig:destiny-hybrid-with-ce} highlights that hybrid calibration can closely align with Bayesian calibration when frequentist error requirements are met automatically at the Bayesian-optimal \(n\). In this example, no sample size reduction can be achieved from hybrid calibration. However, the key reason is that frequentist power at $dp=0.45$ requires a much smaller sample size than a realistic Bayesian design prior centered at $p=0.45$ does. As a consequence, Bayesian power is the limiting factor for hybrid calibration. In other settings, or when a much more informative design prior is selected, this situation might reverse. Then, Bayesian power could require a smaller sample size than frequentist power and a hybrid calibration would reduce the required samples size compared to frequentist calibration. As an example, we increased the informativity in the Bayesian design prior under $H_1$ while keeping it centered at $p=0.45$. 

\begin{figure}[ht]
  \centering
  \includegraphics[width=\textwidth]{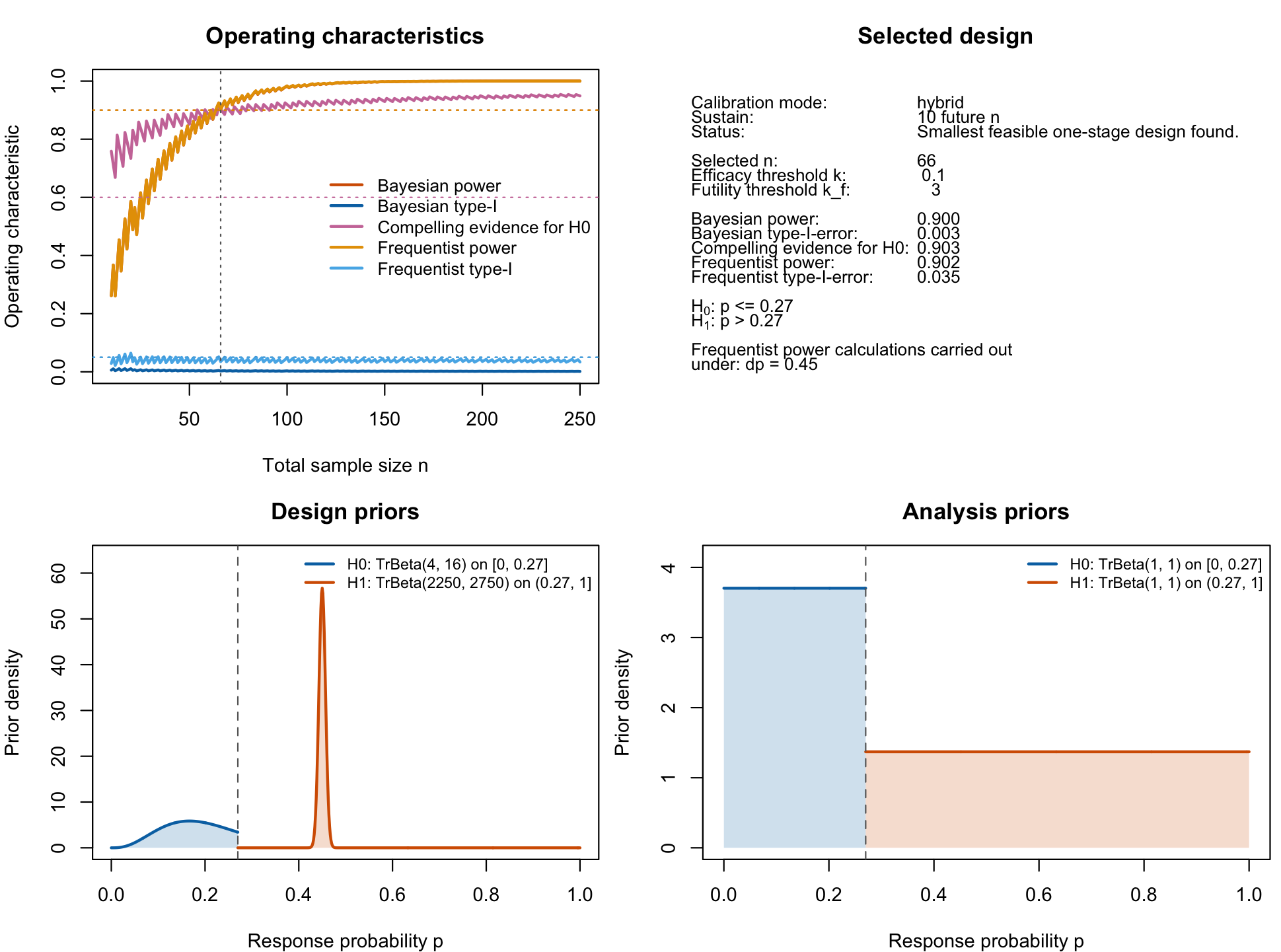}
  \caption{Hybrid calibration for DESTINY-Gastric02 with extremely informative design prior $\mathrm{Beta}(2250,2750)$ under $H_1$. The design simultaneously enforces Bayesian power and frequentist type-I error constraints and leads to a sample size \(n = 66\), corresponding to frequentist calibration results when calculating power under a point-prior at $p=0.45$.}
  \label{fig:destiny-hybrid-with-ce-more-informative}
\end{figure}

\Cref{fig:destiny-hybrid-with-ce-more-informative} shows that the resulting sample size of frequentist calibration (that is, of frequentist power under a point-prior at $p=0.45$) is only reached when increasing the design prior informativity under $H_1$ to a truncated $\mathrm{Beta}(2250,2750)$ prior. This implies, that the informativity in this truncated Beta prior corresponds to having observed $2250+2750=5000$ patients already, out of which $2250$ shows a success and $2750$ a failure. This prior is shown in the bottom left panel in \Cref{fig:destiny-hybrid-with-ce-more-informative} and illustrates how extreme the informativity is. Bayesian and frequentist power now essentially overlap and cannot be separated visually from another.

Only for an \textit{even more} extreme prior, a reduction in sample size compared to frequentist calibration can be obtained, when using Bayesian power. This illustrates how extreme a seemingly harmless power calculation at a point value $p=0.45$ can be, when interpreted from a Bayesian design prior perspective. What is more, the 95\% highest-density-interval of a $\mathrm{Beta}(2250,2750)$ density is $[0.4384,0.4616]$, so from this point of view, investigators should be fairly certain that the drug does work with about $\approx 45\%$. This shows that using a more diffuse design prior is the more realistic choice compared to full frequentist calibration.

\subsubsection{Intermediate summary}
Taken together, these DESTINY-Gastric02 calibrations show how the Bayes-factor approach enables a nuanced compromise between Bayesian and frequentist criteria. Design priors under \(H_0\) and \(H_1\) directly influence the prior-predictive power and \(\mathrm{PCE}(H_0)\), and hence the required sample size, while the calibration mode (Bayesian, frequentist, hybrid) determines which operating characteristics are enforced as constraints. Under the baseline truncated \(\mathrm{Beta}(9, 11)\) prior, the Bayesian calibration demands \(n \approx 217\) to guarantee high power and a large \(\mathrm{PCE}(H_0)\), whereas a more concentrated truncated \(\mathrm{Beta}(24, 30)\) prior under \(H_1\) reduces this requirement to \(n \approx 125\). The frequentist calibration, in turn, yields an even smaller \(n \approx 66\) when only classical power and type-I error are constrained, and the hybrid calibration matches the Bayesian design in this scenario. Increasing the informativity of the design prior under $H_1$ demonstrated how extreme the informativity must be to obtain a sample size as small as the one produced by strictly frequentist calibration with power calculated under $p=0.45$. 

These features make DESTINY-Gastric02 a compelling real-world example of how Bayes-factor based sample size planning can incorporate clinical prior information while maintaining interpretable long-run error control.

\begin{table}[h]
  \centering
  \caption{Comparison of calibrated single-arm Bayes-factor designs for DESTINY-Gastric02 under different calibration modes and $H_1$ design priors. All designs use $p_0 = 0.27$, $p_1 = 0.45$, efficacy threshold $k = 0.1$, and futility threshold $k_f = 3$.}
  \label{tab:destiny-calibration}
  \resizebox{\textwidth}{!}{%
    \begin{tabular}{lccccc}
      \toprule
      Calibration mode &
      $H_1$ design prior &
      $n$ &
      Bayes OC (power / type-I) &
      Freq.\ OC (power / type-I) &
      $\mathrm{PCE}(H_0)$ \\
      \midrule
      Bayesian (baseline) &
      trunc.\ $\mathrm{Beta}(9, 11)$ &
      217 &
      0.901 / 0.002 &
      1.000 / 0.036 &
      0.943 \\[0.5ex]

      Bayesian (concentrated $H_1$) &
      trunc.\ $\mathrm{Beta}(24, 30)$ &
      125 &
      0.906 / 0.002 &
      0.994 / 0.041 &
      0.934 \\[0.5ex]

      Frequentist\textsuperscript{$\dagger$} &
      trunc.\ $\mathrm{Beta}(9, 11)$ &
      66 &
      0.90 / 0.002 &
      0.90 / 0.05 &
      0.903 \\[0.5ex]

      Hybrid\textsuperscript{$\star$} &
      trunc.\ $\mathrm{Beta}(9, 11)$ &
      217 &
      0.901 / 0.002 &
      1.000 / 0.036 &
      0.943 \\
      \bottomrule
    \end{tabular}%
  }

  \vspace{0.5ex}
  \footnotesize\textsuperscript{$\dagger$}For the frequentist calibration, Bayesian power and type-I error are computed but not directly constrained; constraints are imposed on frequentist power and type-I error at $p_0$ and $p_1$. Therefore, the design priors are irrelevant for frequentist calibration, but as shown in the main text, the frequentist calibration results for power correspond to selecting a truncated $\mathrm{Beta}(2250,2750)$ prior on $H_1$.
  
  \footnotesize\textsuperscript{$\star$}For hybrid calibration, Bayesian power and frequentist type-I error are constrained.
\end{table}

\Cref{tab:destiny-calibration} summarizes the results of the different calibration modes and illustrates that while the baseline Bayesian design yields the best operating characteristics, selecting a more optimistic design prior substantially reduces the required sample size for the trial at only a very moderate loss of \(\mathrm{PCE}(H_0)\) and frequentist power. The frequentist type-I-error also increases slightly, but is still controlled. In contrast, strictly frequentist calibration yields a much smaller sample size, but the assumption of the power calculation at $p=0.45$ is questionable from a practical point of view, as it corresponds to an extreme informative design prior under $H_1$.

The question arises which calibration mode to use and why. From a practical point of view, strictly frequentist calibration yields the smallest sample size. Our sample size calculations would have allowed the DESTINY-Gastric02 trial to terminate earlier, after $n=66$ instead of $n=72$ patients, compare \Cref{tab:destiny_gastric02}. However, Bayesian calibration results are much more realistic, and the concentrated design prior which yields $n=125$ patients has a 95\% highest-density-interval $[33.53,55.60]$. This is a much more realistic assumption before observation of any trial data compared to frequentist calibration. In contrast, hybrid calibration or Bayesian calibration with the diffuse $\mathrm{Beta}(9,11)$ prior yields a 95\% highest-density-interval $[0.2739,0.6318]$. As a consequence, smaller and larger success probabilities compared to the point estimate $p=0.45$ are more likely a priori, compared to the more concentrated design prior. This increases sample size, and ultimately, investigators need to answer from a practical domain-specific point of view whether these success probabilities are realistic or not. In the context of the DESTINY-Gastric02 trial, success probabilities as large as $0.6318$ could be unrealistic, favouring the concentrated $\mathrm{Beta}(24,30)$ prior and Bayesian calibration with a sample size of $n=125$.

In closing this example section, note that if from a regulatory perspective, strict frequentist calibration is required, then using the $\mathrm{Beta}(24,30)$ prior and shifting to hybrid calibration solves this requirement. Then, the upper left panel in \Cref{fig:destiny-bayes-conc-with-ce} shows that the required sample size still results in $n=125$.

\subsection{LCMC3: neoadjuvant atezolizumab in resectable NSCLC}

The LCMC3 trial is an open-label, single-arm phase II study of neoadjuvant atezolizumab in patients with resectable non-small cell lung cancer (NSCLC) \citep{chaftNeoadjuvantAtezolizumabResectable2022}. Patients received up to three cycles of atezolizumab before surgery. The primary endpoint was major pathological response (MPR), defined as $\leq 10\%$ viable tumour cells at resection, which is a binary outcome at the patient level. Based on historical experience, an MPR rate around $15\%$ was considered the minimum threshold indicating promising neoadjuvant activity \citep{chaftNeoadjuvantAtezolizumabResectable2022}. The sample size was chosen so that the lower bound of the $95\%$ confidence interval for the MPR rate would exceed $15\%$ if the true rate was in the 25--30\% range, corresponding to $p_0=0.15$ and a clinically interesting target around $p_1\approx 0.25$--$0.30$ \citep{chaftNeoadjuvantAtezolizumabResectable2022}.

In the original analysis, the trial was declared positive if the observed MPR rate met or exceeded the $15\%$ threshold, and the confidence interval excluded rates below $15\%$ \citep{chaftNeoadjuvantAtezolizumabResectable2022}. Within a Bayes-factor design, this can again be framed as a composite null
\[
  H_0: p \leq 0.15
\]
representing uninteresting MPR rates at or below the historical benchmark, and an alternative
\[
  H_1: p > 0.15,
\]
with a design prior on $p$ under $H_1$ centred at $p_1 \approx 0.25$ (or $0.30$) to encode the expected improvement under neoadjuvant atezolizumab. Under $H_0$, a truncated beta prior on $[0,0.15]$ with mean near $0.15$ reflects the belief that higher MPR rates are unlikely without effective immunotherapy, whereas under $H_1$ a truncated beta prior on $(0.15,1]$ with mean around $0.25$--$0.30$ captures the target neoadjuvant effect. This provides a clinically grounded basis for Bayes-factor calibration of power and type-I-error in terms of evidence thresholds on MPR.

\begin{table}[ht]
\centering
\caption{Key design quantities for LCMC3 (single-arm neoadjuvant atezolizumab trial in resectable NSCLC), compare \cite{chaftNeoadjuvantAtezolizumabResectable2022}.}
\label{tab:lcmc3}
\begin{tabular}{ll}
\toprule
Quantity & Value \\
\midrule
Indication & Resectable NSCLC (stages IB--IIIB)  \\
Design & Single-arm phase II, neoadjuvant immunotherapy\\
Primary endpoint & MPR (binary; $\leq 10\%$ viable tumour cells)\\
Benchmark rate $p_0$ & 0.15 (minimum interesting MPR threshold)\\
Target rate $p_1$ & $\approx 0.25$--$0.30$ (anticipated MPR under atezolizumab)\\
Null hypothesis & $H_0: p \leq 0.15$ \\
Alternative hypothesis & $H_1: p > 0.15$ (design prior centred at $p_1\approx 0.25$--$0.30$) \\
Planned sample size & Approximately 140 resected patients (for CI-based design)\\
Decision rule & Lower 95\% CI bound for MPR above 0.15\\
Observed MPR & 20\% (29 of 143; 95\% CI 14--28\%)\\
\bottomrule
\end{tabular}
\end{table}

Table~\ref{tab:lcmc3} summarizes the main design quantities for LCMC3. In addition, we calibrated three one-stage Bayes-factor designs using the \texttt{bfbin2arm} package \citep{kelterTwoArmTwoStage2026}. All calibrations used the composite null $H_0: p \leq 0.15$, the directional alternative $H_1: p > 0.15$, and evidence thresholds on $BF_{01}$ with $k_{\text{ce}} = 3$ for compelling evidence in favour of $H_0$, $k=1/10$ for evidence in favour of efficacy, and target Bayesian power $0.80$ with Bayesian type-I error $0.05$.

First, we adopted a baseline $H_1$ design prior given by a truncated $\mathrm{Beta}(5, 15)$ distribution on $(0.15,1]$, with mean $0.25$ before truncation. Using a stringent efficacy threshold $k = 0.1$, the Bayesian calibration yields a sample size of $n = 295$, with Bayesian power $0.801$, Bayesian type-I error $0.001$, and $\mathrm{PCE}(H_0) = 0.957$; the corresponding frequentist power and type-I error at $p_1 = 0.25$ and $p_0 = 0.15$ are $0.987$ and $0.018$, respectively. This design thus guarantees very strong evidence against $H_0$ when neoadjuvant atezolizumab achieves the anticipated MPR, while also providing a high probability of compelling evidence for $H_0$ if the true MPR is at or below $15\%$.

Second, to illustrate the impact of more informative design priors under $H_1$, we replaced the baseline truncated $\mathrm{Beta}(5, 15)$ prior by a more concentrated truncated $\mathrm{Beta}(40, 120)$ prior with the same mean, but much tighter concentration around $p_1 = 0.25$. Under otherwise identical settings with $k = 0.1$, the calibrated sample size drops to $n = 175$, with Bayesian power $0.808$, Bayesian type-I error $0.001$, and $\mathrm{PCE}(H_0) = 0.937$, and frequentist power and type-I error equal to $0.899$ and $0.018$. This comparison shows that increasing prior informativeness under $H_1$ directly reduces the Bayesianly calibrated sample size, while maintaining essentially the same evidence thresholds and operating characteristics.

Third, we examined the effect of a less stringent efficacy threshold by setting $k = 1/3$, corresponding to requiring only moderate evidence ($BF_{01} \leq 1/3$) against $H_0$. With the concentrated truncated $\mathrm{Beta}(40, 120)$ design prior under $H_1$, the calibrated sample size further decreases to $n = 115$. At this $n$, the Bayesian power is $0.804$, the Bayesian type-I error increases modestly to $0.007$, and $\mathrm{PCE}(H_0)$ remains high at $0.930$, while the frequentist power and type-I error at $p_1 = 0.25$ and $p_0 = 0.15$ are $0.872$ and $0.056$. This design illustrates the trade-off between sample size and strength of evidence: relaxing the Bayes-factor threshold $k$ yields a smaller trial, but at the price of accepting a slightly higher probability of moderate evidence against $H_0$ when the null is true.

Table~\ref{tab:lcmc3-bf-designs} compares these three Bayes-factor designs side by side. Taken together, the LCMC3 example demonstrates how Bayes-factor based sample size planning in a single-arm neoadjuvant setting can incorporate clinically motivated benchmarks, design priors, and evidence thresholds, and how prior informativeness and the choice of $k$ jointly determine the required sample size and the balance between power, type-I error, and the probability of compelling evidence for $H_0$.

\begin{table}[ht]
\centering
\caption{Calibrated Bayes-factor designs for LCMC3 under different $H_1$ design priors and evidence thresholds. All designs use $p_0 = 0.15$, $p_1 = 0.25$, futility threshold $k_{\text{ce}} = 3$, and target Bayesian power $0.80$ with Bayesian type-I error $0.05$.}
\label{tab:lcmc3-bf-designs}
  \resizebox{\textwidth}{!}{%
  \begin{tabular}{lccccc}
\toprule
Calibration & $H_1$ design prior & $n$ & Bayes power / type-I & $\mathrm{PCE}(H_0)$ & Freq.\ power / type-I \\
\midrule
Bayesian (baseline $H_1$, $k = 0.1$) &
trunc.\ $\mathrm{Beta}(5, 15)$ &
295 &
0.801 /
0.001 &
0.957 &
0.987 / 0.018 \\[0.5ex]

Bayesian (concentrated $H_1$, $k = 0.1$) &
trunc.\ $\mathrm{Beta}(40, 120)$ &
175 &
0.808 /
0.001 &
0.937 &
0.899 / 0.018 \\[0.5ex]

Bayesian (concentrated $H_1$, $k = 1/3$) &
trunc.\ $\mathrm{Beta}(40, 120)$ &
115 &
0.804 /
0.007 &
0.930 &
0.872 / 0.056 \\
\bottomrule
\end{tabular}}
\end{table}

\section{Discussion}

This tutorial introduced simulation-free Bayesian power and sample size calculations for Bayes factors in single-arm phase II trials with binary endpoints, using two oncology-motivated examples and the \texttt{bfbin2arm} package \citep{KelterPawel2025}. The key features are:
\begin{itemize}
  \item use of Bayes factors as primary evidence measures,
  \item calibration of Bayesian and frequentist power and type-I-error without Monte Carlo simulation,
  \item explicit targeting of the probability of compelling evidence for the null,
  \item implementation in user-accessible software.
\end{itemize}
These properties align with current calls for innovative trial designs that improve efficiency, ethical conduct, and scientific validity through Bayesian and adaptive methods.

Although the presented approach avoids Monte Carlo simulation, careful specification of the design parameters remains essential. As with any Bayesian design, the choice of prior distribution and the Bayes factor thresholds directly influence operating characteristics and should therefore be justified based on the clinical context and the study objectives.
Future work will extend the examples to three-stage single-arm designs, three-stage two-arm settings, and trials which make use of equivalence testing, all within the same simulation-free calibration framework.

Beyond the methodological contribution, this tutorial aims to lower the barrier to applying Bayesian trial designs in practice. By combining a transparent mathematical framework with an openly available R package, investigators can explore operating characteristics and perform sample size calculations without relying on computationally intensive simulations. 
The proposed framework and accompanying software may facilitate the wider adoption of Bayes-factor based designs in early-phase clinical trials and contribute to more transparent and efficient statistical decision-making.

\appendix

\section{R code for the examples}
\label{app:rcode}

The following R code illustrates how the design in Example~1 can be reproduced with \texttt{bfbin2arm}. The code assumes that the package has been installed from CRAN.\footnote{ee: \url{https://cran.r-project.org/web/packages/bfbin2arm/index.html}.} The full code to reproduce all example results and plots is included in a Quarto file available from the Open Science Foundation.\footnote{Available from the Open Science Foundation repository under \url{https://osf.io/9fjz7/overview?view_only=6ca33c181c25425fa353dd031f5bf6a5}.}

\subsection*{Example 1: The DESTINY-Gastric02 trial investigating second-line T-DXd in HER2-positive gastric cancer}\label{sec:app_ex1}

\begin{verbatim}

> library(bfbin2arm)


## DESTINY-Gastric02: single-arm one-stage BF design


> p0_destiny <- 0.27
> p1_destiny <- 0.45

### Design priors

# Design prior under H0: truncated Beta concentrated 
# near p0 on [0, p0]

> prior_H0_destiny <- list(
>  type  = "trunc_beta",
>  a     = 4,
>  b     = 16,
>  lower = 0,
>  upper = p0_destiny
> )

# Baseline H1 design prior: truncated Beta with mean = 0.45

> prior_H1_destiny_baseline <- list(
>  type  = "trunc_beta",
>  a     = 9,
>  b     = 11,
>  lower = p0_destiny,
>  upper = 1
>

# More concentrated H1 design prior: same mean, larger a+b

> prior_H1_destiny_concentrated <- list(
>  type  = "trunc_beta",
>  a     = 24,
>  b     = 30,
>  lower = p0_destiny,
>  upper = 1
> )

### Evidence thresholds

> k_destiny    <- 1/10  # BF01 <= 0.1 → efficacy
> k_ce_destiny <- 3     # BF01 >= 3   → compelling evidence for H0

### Calibration targets

> target_bayes_power   <- 0.90
> target_bayes_type1   <- 0.05
> target_freq_power    <- 0.90
> target_freq_type1    <- 0.05

### Pure Bayesian calibration

# Under a diffuse design prior under $H_1$:

> design_destiny_bayes_baseline <- design_singlearm_onestage_bf(
>   n_min       = 10,
>   n_max       = 250,
>   k           = k_destiny,
>   k_ce        = k_ce_destiny,
>   p0          = p0_destiny,
>   type        = "direction",
>   dp          = p1_destiny,
>   a0          = 1,  b0 = 1,
>   a1          = 1,  b1 = 1,
>   da0         = prior_H0_destiny$a,
>   db0         = prior_H0_destiny$b,
>   da1         = prior_H1_destiny_baseline$a,
>   db1         = prior_H1_destiny_baseline$b,
>   calibration     = "Bayesian",
>   target_power    = target_bayes_power,
>   target_type1    = target_bayes_type1,
>   compute_freq_oc = TRUE
> )

> print(design_destiny_bayes_baseline)
> summary(design_destiny_bayes_baseline)
> plot(design_destiny_bayes_baseline)
\end{verbatim}

\section*{Acknowledgements}
Funded by the Deutsche Forschungsgemeinschaft (DFG, German Research Foundation) – Project number 549296018.

\bibliography{library}

\end{document}